\documentclass{PoS}
\usepackage{amsmath}
\usepackage{amssymb}
\usepackage{subfigure}

\title{Calculating the two-pion decay and mixing of neutral K mesons }

\ShortTitle{Calculating the two-pion decay and mixing of neutral K mesons }

\author{\speaker{Norman Christ}%
         \thanks{This work was supported in part by US DOE grant DE-FG02-92ER40699}\\
        Department of Physics, 538 W120$^{th}$ Street, Columbia University, New York, NY, 10027, USA\\
        E-mail: \email{nhc@phys.columbia.edu}}

\author{RBC and UKQCD collaborations}

\abstract{The recent calculation of the complex isospin-two decay amplitude $A_2$ 
with physical kinematics is presented together with exploratory
calculations of the isospin-zero decay amplitude $A_0$.  Prospects for 
accurate calculation of $A_0$ as well as the mass difference between 
the $K_L$ and $K_S$ mesons are discussed.}

\FullConference{The 30th International Symposium on Lattice Field Theory\\
		 June 24 - 29,  2012\\
		 Cairns, Australia}

\begin{document}

\section{Introduction}

Advances in both computer capabilities and numerical methods now make it possible to study physically light pions in large spatial volumes using a chiral lattice fermion formulation.  This allows the study of standard quantities such as particle masses and decay constants with enhanced precision, removing the uncertainties associated with using chiral perturbation theory to extrapolate to physical quark masses.  This ability to work directly with physical quark masses also allows us to tackle the calculation of more complex quantities where the effects of using unphysically large quark masses may be more difficult to estimate.  In this paper we will discuss the calculation of three such quantities, the decay amplitudes $A_0$ and $A_2$ for the decay of a K meson into $I=0$ and $I=2$ two-pion final states and the mass difference, $\Delta M_K$ between the $K_L$ and $K_S$ neutral kaons.

For $K\to\pi\pi$ decay we will discuss the calculation of the $\Delta I=3/2$ amplitude, $A_2$, with the kaon and pion masses and the pion relative momenta taking their physical values~\cite{Blum:2011ng,Blum:2012uk}.  By working at physical kinematics, the dominant errors come from computational issues such as non-zero lattice spacing and finite volume which can be reduced in future calculations by simply working at smaller lattice spacing and larger volume. We do not need to deal with an uncertain theoretical framework to correct for the absence of the physical $\pi-\pi$ relative momentum or unphysical pion or kaon masses.  For the more difficult calculation of the $\Delta I=1/2$ amplitude, $A_0$ and $\Delta M_K$, where such a physical calculation is not yet possible, we concentrate on developing the computational methods which should allow a calculation with physical kinematics in the not-too-distant future.  

\section{$K\to\pi\pi$ decay}

Since the weak interaction $W^\pm$ bosons which mediate this decay are far too massive to simulate in a lattice QCD calculation, our first step must be to represent $W^\pm$ exchange by the effective four-Fermi interaction which results if the $W$ exchange process is treated as taking place at a space-time point.  The resulting effective Hamiltonian is written as
\begin{align}
 \mathcal{H}^{(\Delta S=1)} &= 
 	\frac{G_F}{\sqrt{2}} V_{ud} V^*_{us} 
		\left\{ \sum_{i=1}^{10} \left[ z_i(\mu)
                   - \frac{V_{td}}{V_{us}^*} \frac{V_{ts}^*}{V_{ud}} y_i(\mu) \right] Q_i \right\}.
\label{eq:Hweak}
\end{align}
Here $V_{qq'}$ is the Cabibbo-Kobayashi-Maskawa matrix  element connecting the charge $-1/3$ quark $q'$ to the charge $+2/3$ quark $q$.  The Wilson coefficient functions $y_i(\mu)$ and $z_i(\mu)$ depend on the scale $\mu$ at which the four-quark operators $Q_i$ are normalized and have been determined in QCD and electro-weak perturbation theory through second order; see Ref.~\cite{Buchalla:1995vs}  for a thorough discussion.  

The ten dimension-six, four-quark operators $Q_i$, $1 \le i \le 10$ are not independent but arise from particular phenomena and are defined in Eqs.~(4-23) of Ref.~\cite{Blum:2001xb}.  The operators $Q_i$, $i=1,2$ represent the naive current-current interaction resulting from simple $W^\pm$ exchange and transform under $SU_L(3) \times SU_R(3)$ as both $(8,1)$ and $(27,1)$ representations.  Those with $i=3$ through 6 result from QCD penguin graphs in which a quark-anti-quark pair emerges from the point-like $W$ exchange and annihilate into a gluon which subsequently creates a possibly different $q-\overline{q}$ pair.  These operators transform in the $(8,1)$ representation.  The final four operators arise from electro-weak penguin diagrams in which a quark-anti-quark pair emerging from the point-like $W$ exchange annihilate to create a photon or $Z$ boson and transform as $(8,8)$, $(8,1)$ and $(27,1)$.  While one order smaller in electro-weak perturbation theory, the $(8,8)$ operators can be important because of the suppression of the leading order, $\Delta I=3/2$ component of the usual $(27,1)$ matrix elements which arises from the ``$\Delta I=1/2$ rule" and an extra power of $M_K^2$ appearing in chiral perturbation theory.  

While this historical classification of the ten $Q_i$ operators is convenient for discussing the electro-weak phenomena underlying the decay, when discussing renormalization and operator mixing it is better to work with a set of seven linearly independent operators, four transforming in the $(8,1)$ representation, one in the $(27,1)$ and two transforming in the $(8,8)$ representation.  If renormalization conditions are imposed in which the flavor-symmetry breaking quark masses can be neglected then these three classes of operators will renormalize independently.  

Critical to an accurate lattice calculation of a weak decay process such as $K\to\pi\pi$ is the ability to relate the lattice-regulated operators appearing in such a calculation to the continuum operators for which the Wilson coefficients were originally computed.  This can be done with increasing precision using the intermediate Rome-Southampton RI/MOM approach~\cite{Martinelli:1995ty}, enhanced by a number of refinements over the past decade.  Here one introduces a regularization-independent scheme to normalize the lattice operators in which particular Landau-gauge-fixed, spin-color-projected, Green's functions are normalized at large, off-shell momenta at a scale characterized by $\mu$.  In this way, the use of lattice perturbation theory is avoided and the original bare lattice operators are expressed in terms of operators renormalized in a scheme which has a well-defined continuum limit.  Because of this non-perturbative step, these methods are referred to as non-perturbative renormalization (NPR).  If $\mu$ is sufficiently large, these same conditions can be accurately imposed in a perturbative calculation allowing this RI/MOM renormalization to be connected to the standard perturbative $\overline{\rm MS}$ continuum scheme in which the Wilson coefficients are typically evaluated.

The effectiveness of these NPR techniques can be seen in the size of the normalization errors presented below.  Three important developments, all used in the calculations discussed here, have made these NPR techniques more accurate.  The first is the use of non-exceptional momenta when imposing the RI/MOM normalization conditions~\cite{Aoki:2007xm}.  This choice of momenta makes these normalization conditions infrared safe, significantly reducing the contributions of small loop momenta and making the use of QCD perturbation theory to relate the RI/MOM and  $\overline{\rm MS}$ schemes more reliable.  The second improvement recognizes that even at $O(a^2)$, lattice artifacts appear in the RI/MOM normalization conditions which are not $O(4)$ invariant.  As a result the amplitudes being studied depend not only on the usual Lorentz scalars such as $p^2$ or $p\cdot p'$, where $p$ and $p'$ are lattice momenta, but also on the direction of these momenta relative to the 4-D lattice axes.  This few-percent direction dependence introduces irregularities into what should be smooth dependence on $p^2$ and prevents the evaluation of the continuum limit if an $a^2$ extrapolation is attempted using lattice momenta in different directions and consequently with different $a^2$ corrections.  These difficulties can be avoided by imposing twisted boundary conditions on the fermion propagators used to evaluate the RI/MOM normalization conditions.  By varying the degree of twist (the fermion phase change when passing through a boundary) the magnitude of the lattice momentum can be varied, {\it e.g.} to compensate for a change in lattice spacing when extrapolating to the continuum limit, without changing the direction of that momentum relative to the underlying lattice~\cite{Arthur:2010ht, Arthur:2011cn}.

The third advance in NPR methods is using the momentum dependence of the RI/MOM renormalization factors to relate operators renormalized at significantly different scales~\cite{Arthur:2010ht, Arthur:2011cn}.  This allows operators which are evaluated in a coarse lattice calculation, such as the $K\to\pi\pi$ amplitudes discussed below to be accurately renormalized.  In this $K\to\pi\pi$ calculation the physical size of the lattice is increased by using a relatively large lattice spacing of 0.144 fm.  This limits the size of the external momenta that can be employed in an RI/MOM renormalization condition to $\lesssim 1$ GeV, a scale too low for the application of perturbation theory.  However, by using the step-scaling function determined in a companion calculation on a finer lattice, these operators renormalized at 1 GeV can be accurately renormalized also at 3 GeV, a scale at which a perturbative calculation can be used to convert to the standard $\overline{\rm MS}$ scheme.

The final difficulty that must be overcome to carry out a realistic lattice calculation of $K\to\pi\pi$ decay arises from the non-zero relative momentum of the physical final state pions.  In the usual lattice QCD calculation, one uses the Euclidean time-development operator $e^{-Ht}$ at large $t$ to project onto the QCD energy eigenstates with the lowest energies.  While this strategy works well to construct the initial $K$ meson, when applied to a state of two pions, unphysical, threshold states with zero relative momentum pions result.

This difficulty can be overcome in two steps.  The first, proposed by Lellouch and Luscher~\cite{Lellouch:2000pv}, recognizes that in finite volume there will be additional excited states containing two pions obeying energy quantization conditions that depend on the volume.  With an appropriately chosen volume, the resulting two-pion energy can be adjusted to equal that of the decaying kaon.  Thus, by identifying the transition amplitude to this excited state and introducing a finite-volume correction factor derived in Ref.~\cite{Lellouch:2000pv} the physical, on-shell, $K\to\pi\pi$ decay amplitude can be computed with controlled errors.  However, the extraction of such an excited state is typically difficult and a second step of imposing boundary conditions to remove some or all of the energy-non-conserving, lower energy $\pi-\pi$ states from the calculation can substantially improve the result~\cite{Kim:2002np, Kim:2003xt, Kim:2009fe}.  We will discuss such boundary conditions in greater detail below.

\subsection{$K\to\pi\pi$ decay with $\Delta I = 3/2$}

The $K\to\pi\pi$ decay amplitude most accessible to the methods of lattice QCD is $A_2$ which describes the  decay into the $\pi-\pi$ state with isospin 2.  This state does not have vacuum quantum numbers and quark flavor conservation implies that the valence quark lines connect the initial kaon, the effective four-Fermi weak operator and the final two pions. Thus, there are no disconnected diagrams, no need for a vacuum subtraction and the light sea quarks do not play a critical role, allowing boundary conditions to be applied only to the valence quarks with resulting errors that vanish exponentially in the lattice size~\cite{Sachrajda:2004mi}.

We ensure that the $I=2$, $\pi-\pi$ final state in a physical $K$ decay is also the lowest energy $I=2$ state in our lattice calculation by imposing anti-periodic boundary conditions on one of the quarks in each pion and adjusting the lattice volume so that $\pi/L$ is close to the 205 MeV momentum of the physical decay pions.  This is accomplished in two steps.  First we use isospin symmetry to relate a physical $\Delta I=3/2$, $K$ decay amplitude to a transition caused by a related weak operator carrying charge $+1$: $K^+\to\pi^+\pi^+$.  We then impose anti-periodic boundary conditions on the $d$ quark.  The  resulting $K^+$ meson will obey periodic boundary condition so the initial $K^+$ can be given zero momentum.  However, each $\pi^+$ mesons will have non-zero momentum equal to $\pi/L$ except for rescattering effects.  Normally the use of such isospin violating boundary conditions would introduce potentially dangerous mixing between the $I=2$ final state of interest and the $I=0$ state which overlaps with the vacuum.  However, by arranging the final state to carry charge 2 we have assured that no $I=0$ final state is possible.

The first calculation adopting this strategy was C.~Kim's Ph.D. thesis~\cite{Kim:2004sk} which used a quenched $16^3\times 32$ ensemble and unphysically massive kaons and pions.  An interesting related calculation~\cite{Yamazaki:2008hg} studied the decay of a kaon carrying non-zero momentum, an alternative technique to achieve a final $\pi-\pi$ final state with physical relative momenta. The statistical errors and resulting practical restriction on the separation between the kaon and $\pi-\pi$ sources in this non-zero momentum approach appear to favor the boundary condition strategy discussed here.

\begin{figure}[ht]
\centering
\subfigure[The setup for the calculation of $A_2$ with the kaon wall source at the time the $t_K$, the weak operator at $t$ and the two-pion source the time $t_\pi$.]
{\includegraphics[width=0.42\textwidth]{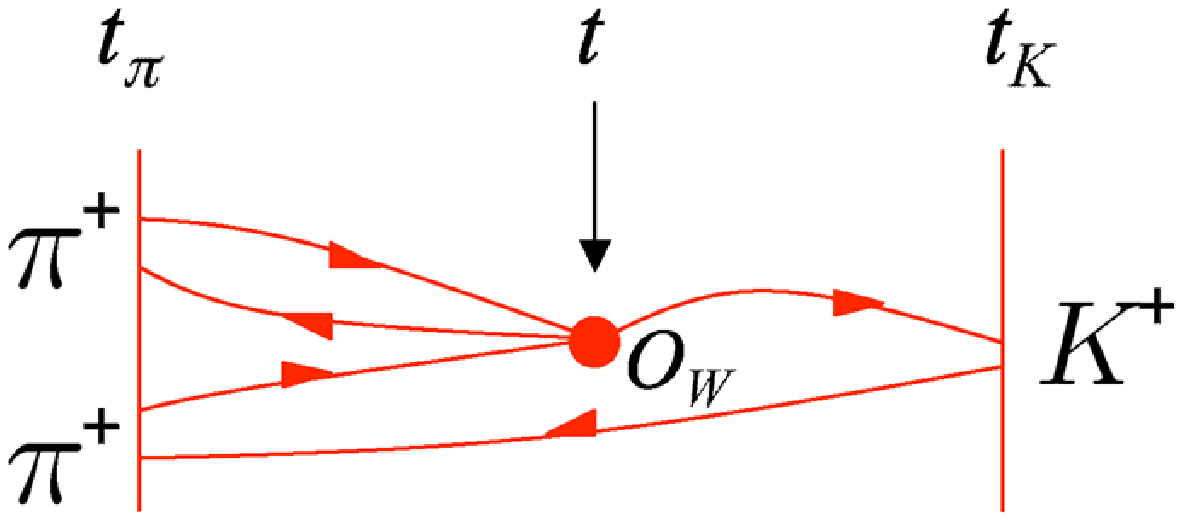}
\label{fig:CompSetUp}} \quad
\subfigure[\label{sfig:ps1} Comparison of calculated phase shift with experimental results \cite{Hoogland:1977kt, Losty:1973et} and a phenomenological curve~\cite{Schenk:1991xe}.]
{\includegraphics*[width=0.35\textwidth]{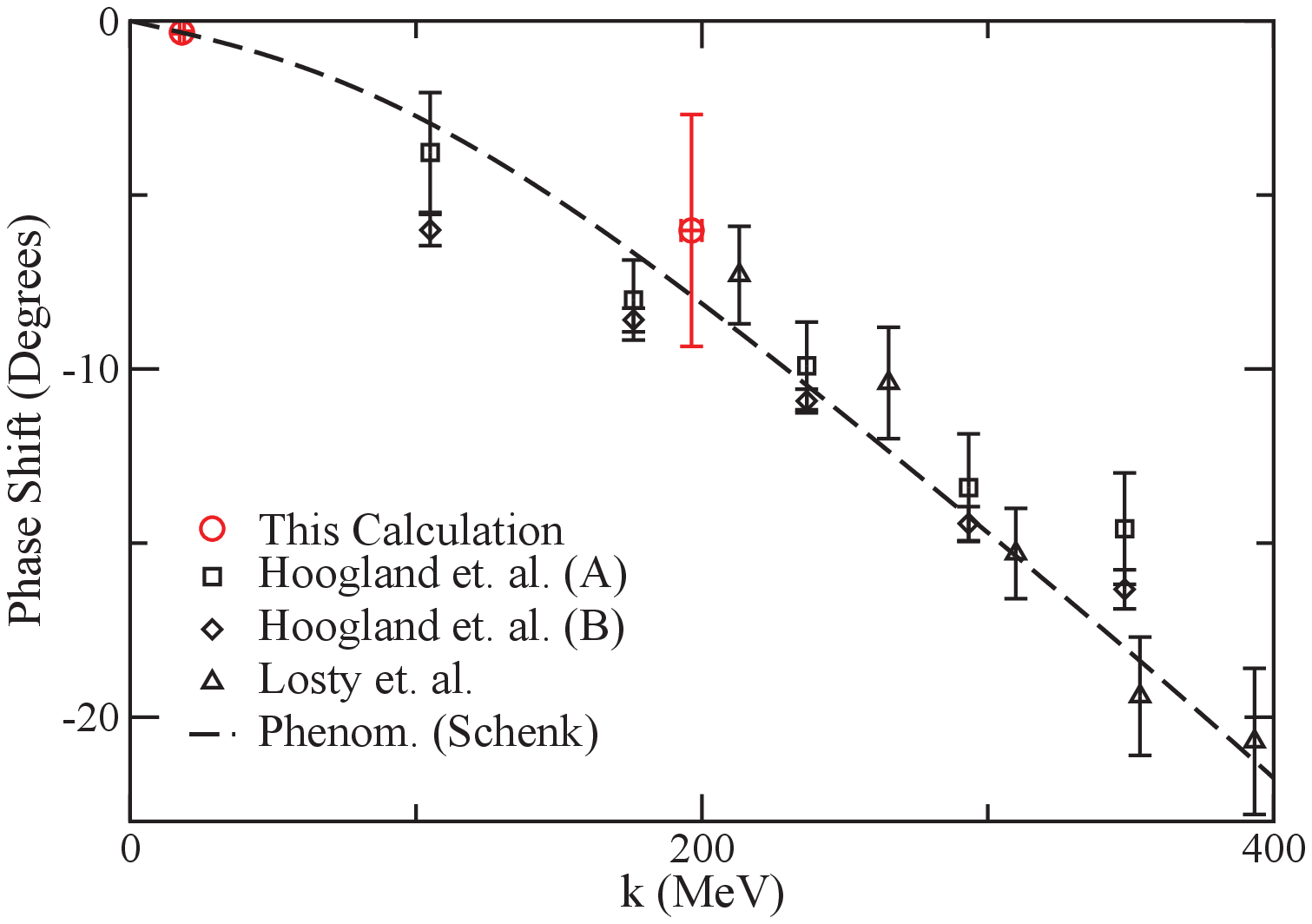}} 
\label{fig:delta_pi-pi}
\caption{}
\end{figure}

We now turn to the recent RBC/UKQCD calculation of $A_2$, using a large lattice spacing of $1/a=1.364(9)$ GeV, a near physical (partially quenched) pion mass of 142.11(94) MeV, a kaon mass of 505.5(3.4) MeV and a two-pion energy of 484.5 (4.2) MeV~\cite{Blum:2011ng, Blum:2012uk}.  This two-pion energy corresponds to a $d$ quark which obeys anti-periodic boundary conditions in two of the possible three spatial directions.  Three independent effective weak operators contribute to the complex amplitude $A_2$: $Q^{(27,1)}$, $Q^{(8,8)}$ and $Q^{(8,8)m}$.  The computational setup is shown schematically Fig.~\ref{fig:CompSetUp}. The light and strange quark propagators are computed using both periodic and anti-periodic in the time direction.
The strange quark source is located at the $t_K$, where seven values of $t_K$ are used varying between $t_\pi+20$ and $t_\pi+44$ in steps of four time units where $t_\pi$ locates of the source of the light quarks.   The sum of the periodic and anti-periodic propagators are used for both the light and strange quarks to increase statistics and reduce around-the-world effects.

\begin{figure}[t]
\centering
\subfigure[$Q^{(27,1)}$]{\includegraphics*[width=0.32\textwidth]{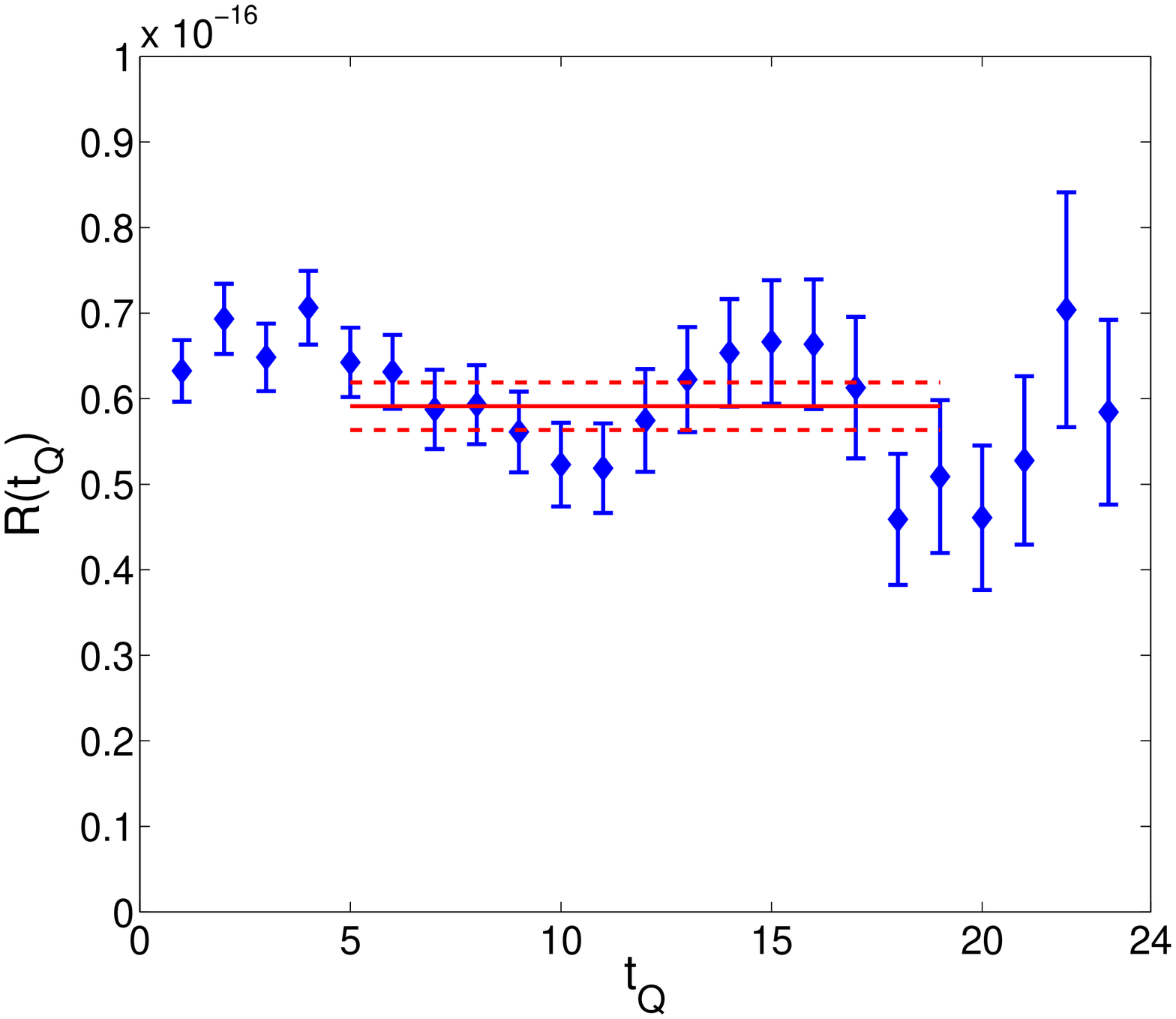}}
\subfigure[$Q^{(8,8)}$]{\includegraphics*[width=0.32\textwidth]{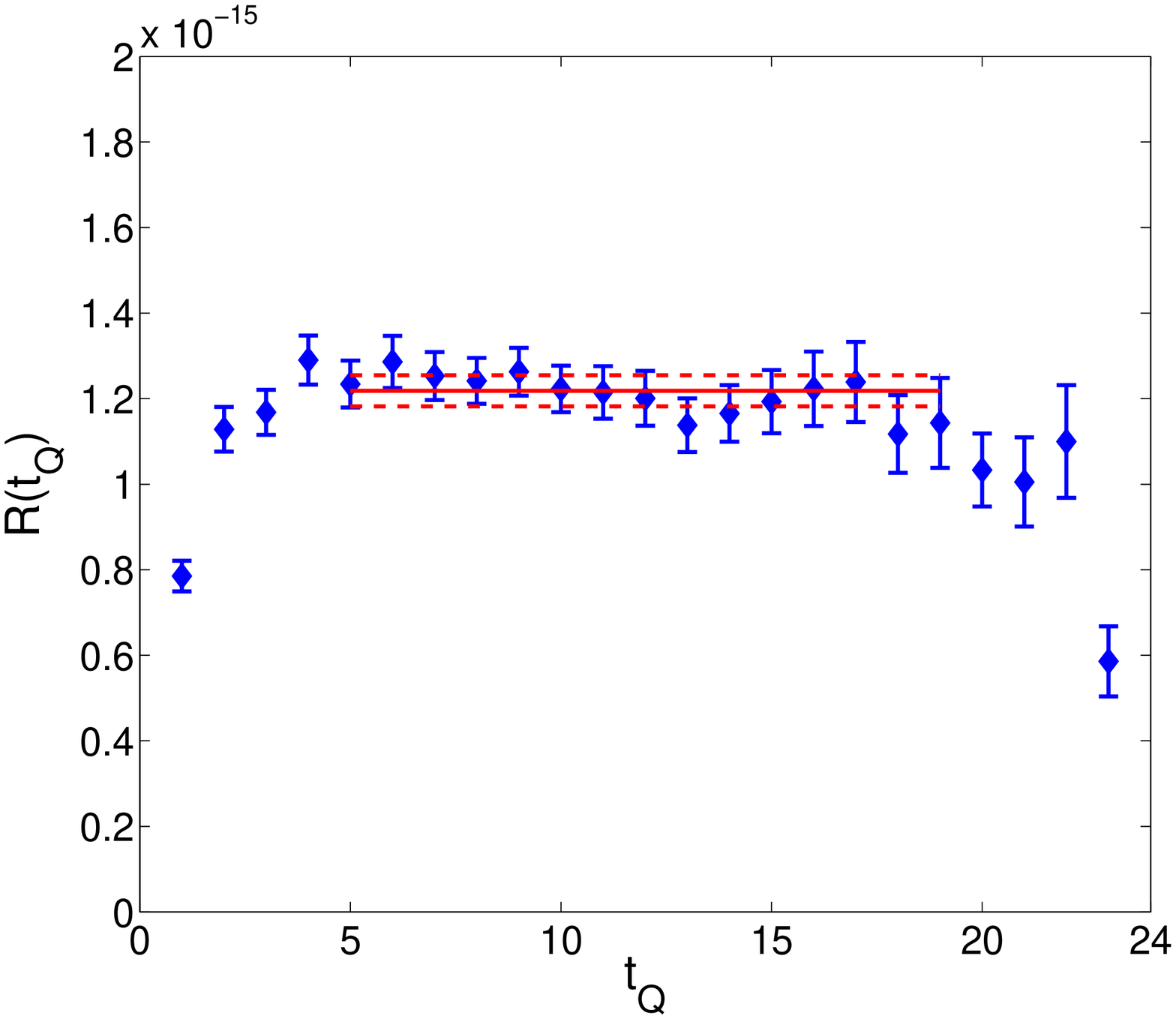}}
\subfigure[$Q^{(8,8)m}$]{\includegraphics*[width=0.32\textwidth]{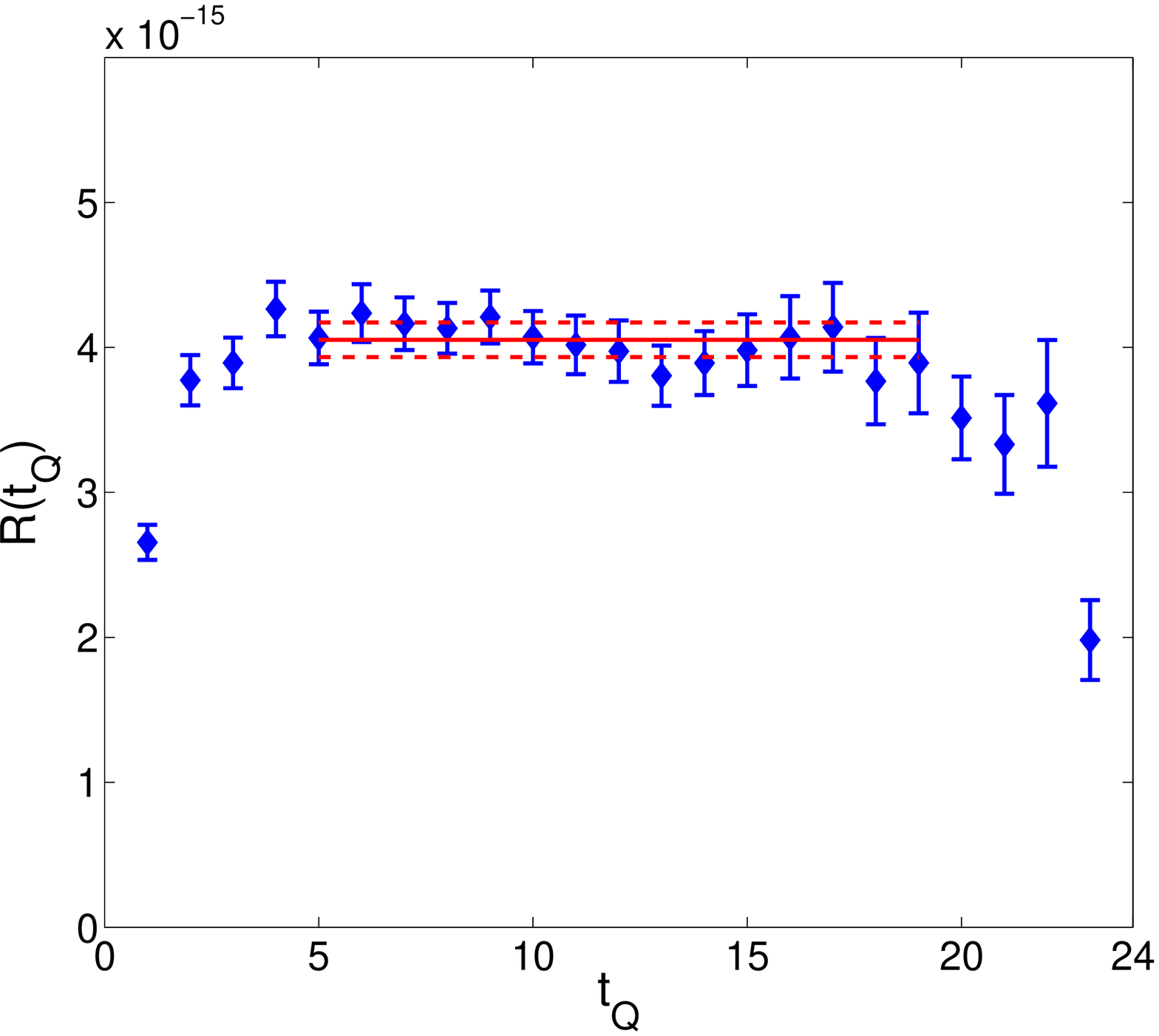}}
\caption{\label{fig:A2_data} Ratios of three point and two point functions described in the text from which the matrix elements of the three operators $Q^{(27,1)}$ , $Q^{(8,8)}$ and $Q^{(8,8)m}$ are determined, plotted as a function of the time $t_Q$ at which the operator is located.  These are computed using anti-periodic boundary conditions in two spatial directions for our full ensemble of 146 configurations.   The two-pion source is located at $t_\pi=0$ and the kaon source at $t_K=24$.}
\end{figure}

In Fig.~\ref{fig:A2_data} we show a ratio $R^i(t_Q)$ of three-point to two-point correlation functions from which the lattice matrix elements of the three operators $Q^{(27,1)}$ , $Q^{(8,8)}$ and $Q^{(8,8)m}$ are determined:
\begin{equation}
R^i(t_Q) = \frac{C^i_{K\pi\pi}(t_Q)}{C_K(t_K-t_Q)C_{\pi\pi}(t_Q)} = \frac{\mathcal{M}_i}{Z_K Z_{\pi\pi}}.
\end{equation}
Here the location of the two-pion source has been set to $t_\pi=0$ and the label $i$ distinguishes the three $\Delta I=3/2$ operators being studied.  The amplitudes $\mathcal{M}_i$ are the lattice matrix elements we are trying to determine while $Z_K$ and $Z_{\pi\pi}$ are source normalization factors which can be determined directly from the two-point functions. The ratio $R^i(t_Q)$ should not depend on $t_Q$ if only kaon and two-pion states are present.  The solid line shown in each graph indicates the fitted results, the dotted lines the width of the error band and the horizontal position of these lines, the range over which the fit is performed.

The complex amplitude $A_2$ can be obtained from the lattice matrix elements $\mathcal{M}_i$ using:
\begin{equation}
A_2 = \frac{1}{a^3}\left[\frac{1}{2\pi q_\pi}
     \sqrt{\frac{\partial\phi}{\partial q_\pi}+\frac{\partial\delta}{\partial q_\pi} } \right]\sqrt{\frac{3}{2}}\frac{G_F}{2\pi} V_{ud} V_{us}^* 
               \sum_{i,j} C_i(\mu)Z_{ij} m_K^{3/2}\mathcal{M}_j,
\label{eq:A_2_master}
\end{equation}
obtained from a combination of Eqs.~(18) and (20) in Ref.~\cite{Blum:2012uk}.  The Wilison coefficients $C_i(\mu)$, evaluated at $\mu=3$ GeV are obtained from next-leading-order formulae in Ref.~\cite{Buchalla:1995vs}.  The renormalization matrix, $Z_{ij}$, transforms the lattice operators used on the DSDR ensemble into the $\overline{\rm MS}$ scheme at $\mu = 3$ GeV. This matrix is determined using the step scaling methods described above.

The square brackets contain the Lellouch-Luscher correction factor with its sum of derivatives of a known kinematic function $\phi$ and the $I=2$ $\pi-\pi$ $s$-wave phase shift $\delta$.  This phase shift can be determined directly by applying the Luscher quantization condition~\cite{Luscher:1990ux} to the calculated $\pi-\pi$ energies for zero and two twists.  The results are shown as the two red open circles in Fig.~\ref{fig:delta_pi-pi} and contribute only 6\% to the factor in square brackets.

Table~\ref{tab:syserrors} lists the systematic errors in this calculation.  The largest comes from the large finite lattice spacing and is estimated in two ways.  First we vary the physical quantity that is used to determine the lattice scale, using $m_\Omega$, $f_\pi$, $f_K$ and $r_0$.  Since the lattice spacing enters Eq.~\eqref{eq:A_2_master} with the third power, this explicit, $\approx 5\%$ uncertainty is amplified $3\times$.  We obtain a similar estimate from the $a^2$ dependence of the $K^0-\overline{K^0}$ matrix element of the related (27,1) operator which determines $B_K$ and has been computed at a number of lattice spacings.  A second error that should be mentioned arises from partial quenching of the light quark mass, the valence and dynamical pions having masses of 142 MeV and 171 MeV respectively.  We estimate an upper bound on the resulting error from threshold calculations of $A_2$ performed on the RBC/UKQCD $1/a=2.28$ GeV ensembles in which the dependence on the sea quark mass was zero within statistical errors.

\begin{table}[t]
\begin{center}
\begin{tabular}{l|c|c}
& $\textrm{Re}A_2$ & Im$A_2$ \\
\hline \hline
lattice artifacts& 15\% & 15\% \\
finite-volume corrections& 6.0\%& 6.5\%\\
partial quenching & 3.5\%& 1.7\%\\
renormalization & 1.8\%& 5.6\%\\
unphysical kinematics & 0.4\%& 0.8\%\\
derivative of the phase shift&0.97\%& 0.97\%\\
Wilson coefficients &6.6\%& 6.6\% \\
\hline
Total & 18\%& 19\%  \\
\end{tabular}
\caption{Estimates of the major systematic errors in this calculation of Re\,$A_2$ and Im\,$A_2$. \label{tab:syserrors}}
\end{center}\end{table}

We find Re$(A_2)=1.381(46)_{\textrm{stat}}(258)_{\textrm{syst}}\,10^{-8}$ GeV and Im$(A_2)=-6.54(46)_{\textrm{stat}}(120)\,_{\textrm{syst}}10^{-13}$ GeV.  Here Re\,$A_2$ agrees well with the experimental values of $1.479(4)$ and $1.573(57)\,10^{-8}$\,GeV obtained from $K^+$ and $K_S$ decays respectively.  The difference between these two experimental numbers results from isospin breaking effects, which are not included in our calculation. The imaginary part of $A_2$ is unknown so that this result represents its first direct determination. 

\subsection{$K\to\pi\pi$ decay with $\Delta I = 1/2$}

\begin{figure}[t]
\centering
\includegraphics*[width=0.5\textwidth]{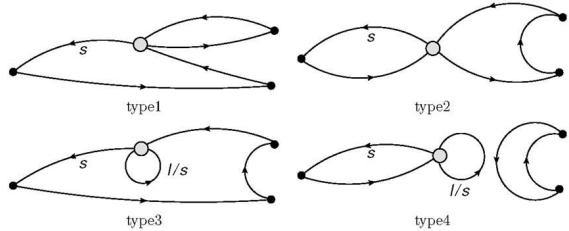}
\caption{\label{fig:A0_types} The topologies distinguishing the four types of diagram contributing to the $I=0$ amplitude $A_0$}
\end{figure}

The calculation of the  $\Delta I = 1/2$ amplitude $A_0$ describing kaon decay into the $I=0$, $\pi-\pi$ state is much more difficult than that for $A_2$.  A total of 50 different contractions contribute which can be organized into the four types shown in Fig.~\ref{fig:A0_types}.  The greatest difficulty is cause by disconnected diagrams shown as type 4.  Such diagrams lead to a signal-to-noise ratio which decreases exponentially with increasing time separation and imply that both sea and valence quarks enter in the physical propagating states.  This requires that if boundary conditions are used to remove unwanted zero relative momentum $\pi-\pi$ states, these conditions must be imposed both when computing the valance propagators and when generating the gauge ensembles.  A further difficult comes from the quadratically divergent quark loops found in diagrams of type 3 and type 4.  While these terms do not contribute to on-shell matrix elements, they can enhance off-shell, excited state contributions by factors of 10-20 and some partial subtraction must be carried out if the usual large-time methods are to be able to successfully remove the resulting excited state contamination. 

However, while severe, these difficulties may be more easily overcome for a quantity such as $A_0$ which involves light pions with their positive definite propagators and a kaon, which has half the mass of the much more difficult nucleon.  Here we will summarize recent results for $A_0$ obtained with unphysical, threshold kinematics and relatively heavy pions which appear in Ref.~\cite{Blum:2011pu} and in the Ph.D. thesis of Q.~Liu~\cite{Liu:2012xx}.  The former were obtained from 800 configurations using 2+1 flavors, $1/a=1.73$ GeV and an $16^3\times32$ lattice volume.  In contrast to the calculations of $A_2$ described earlier in which the two-pion was fixed on a single time slice, the correlation functions used in the calculation of $A_0$ were computed for each of the possible 32 time slices.  With these large statistics, it was possible for the first time to determine Re$(A_0)$ from an explicit $K\to\pi\pi$ matrix element, with a statistical error of $\approx 25$\%. Of course, this 25\% error was achieved for unphysical threshold kinematics in which the two pions are at rest and for an unphysically large pion mass of 422 MeV.  (The valence strange quark mass was adjusted to make $m_K=E_{\pi\pi}$ so that an energy conserving decay was studied.)  The imaginary part of $A_0$ could not be distinguished from noise.  This quantity is dominated by QCD penguin diagrams of type 3 and is more difficult to compute.

\begin{figure}[t]
\centering
\subfigure[$Q_2$, which gives the dominant contribution to Re$(A_0)$]{\includegraphics*[width=0.45\textwidth]{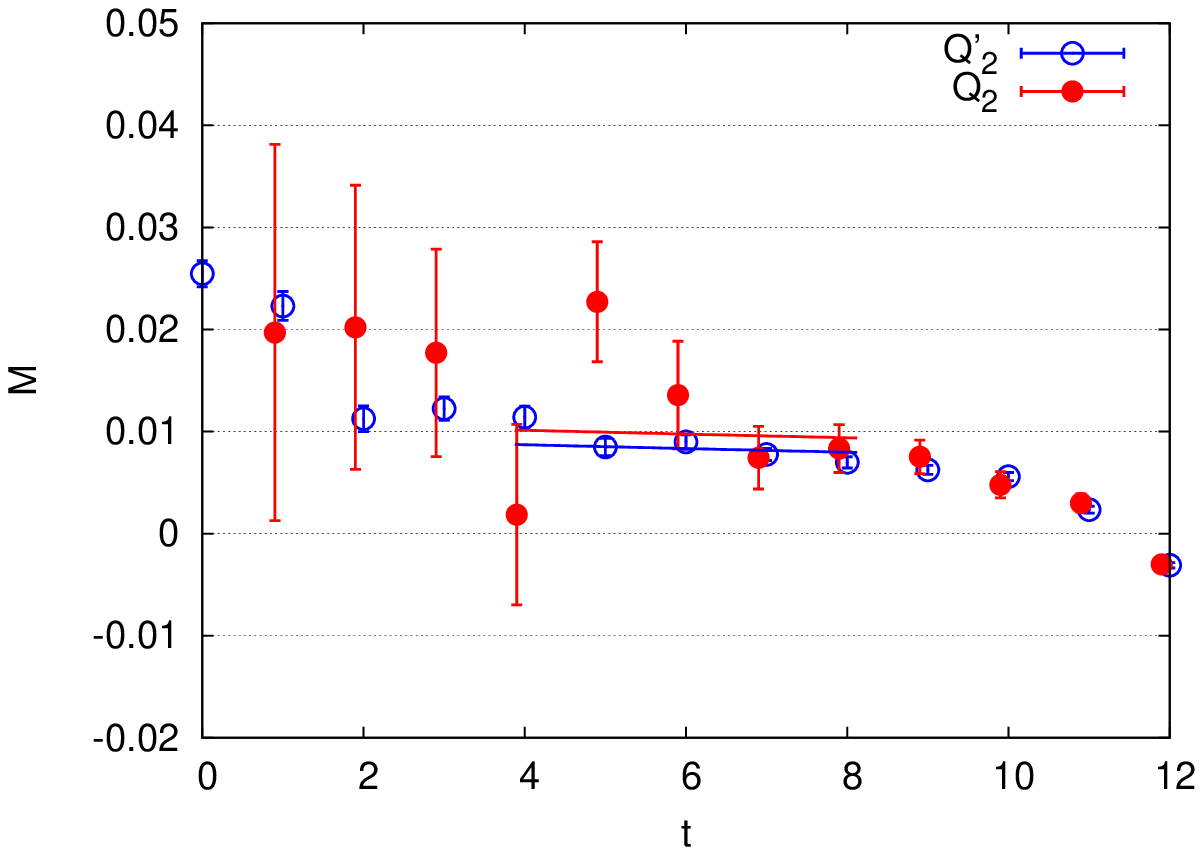}}
\subfigure[$Q_6$, which the dominant contribution to Im$(A_0)$]{\includegraphics*[width=0.45\textwidth]{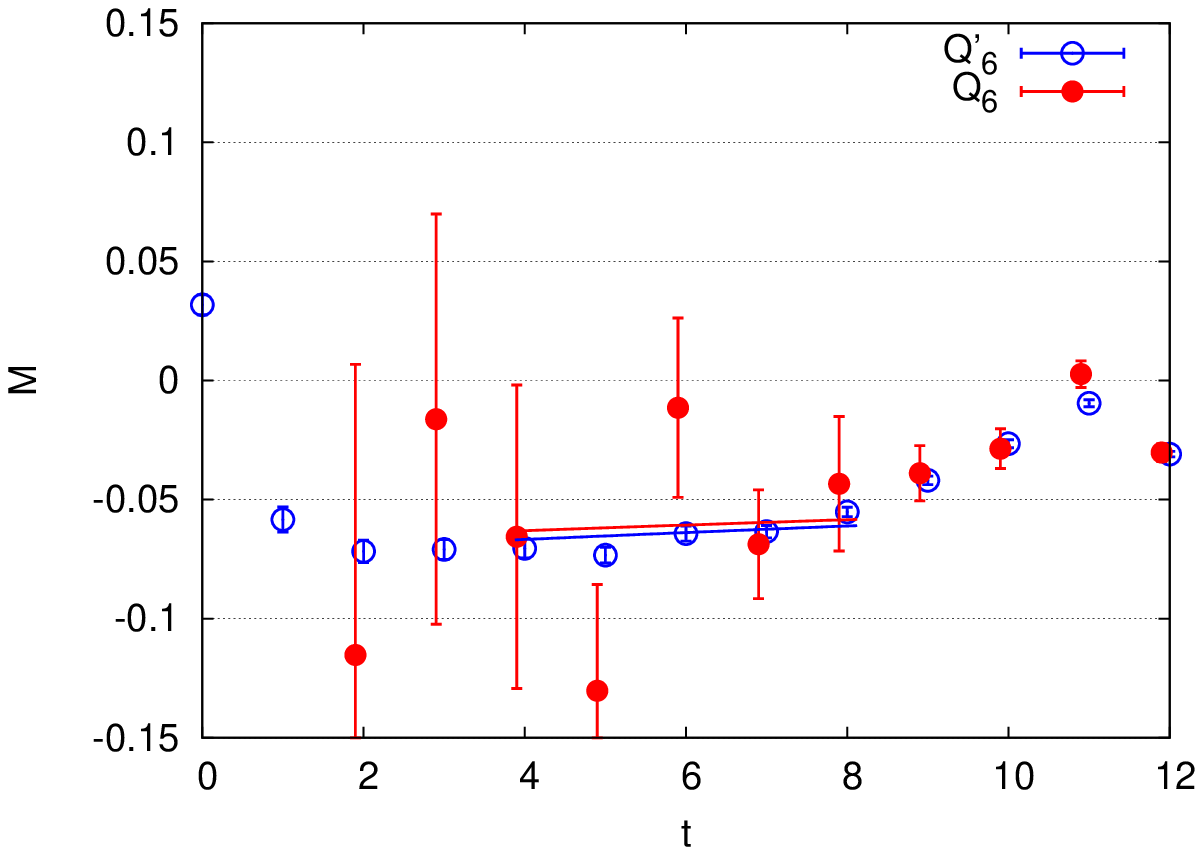}}
\caption{\label{fig:A0_24} The point functions for the operators $Q_2$ and $Q_6$ as a functino of $t$ the time slice over which that operator is summed.  In this figure, the pions in the two-pion source are located at the times $-4$ and $0$ and the kaon source at time 12.  The close points show the result if the disconnected diagrams are omitted while the open point show the complete result.}
\end{figure}

Exploiting what was learned in this $16^3\times32$ calculation, a more ambitious $24^3\times64$ calculation was undertaken~\cite{Liu:2012xx}, again using the same gauge action and a smaller light quark mass, giving a 329 MeV pion mass.  Two important improvements were realized.  First, propagators both periodic and anti-periodic in the time were evaluated.  By using the sum of these propagators, the distance to the nearest periodic image of the source is moved from 64 to 128 time units, substantially reducing the around-the-world effects which gave the largest excited state contamination from the divergent type 3 diagrams in the $16^3\times32$ calculation.  Second, the two-pion source was modified so that the pions were emitted from different time slices.  (We refer to this as a split-pion source.)  This substantially reduced the coupling to the vacuum state and the resulting noise in the disconnected diagrams.  With these advances, improved results were obtain using only 137 instead of 800 configurations.  Now both the real and imaginary parts of $A_0$ can be resolved.   The three-point functions for the two dominant matrix elements are shown in Fig.~\ref{fig:A0_24}.  The open symbols show results without the disconnected graphs while the closed symbols show the full result.  While the statistical errors are much larger for the full amplitudes, the central values are consistent with those coming from only connected graphs.  At present the disconnected diagrams are only a source of noise. 

With the demonstration of methods capable of resolving both the real and imaginary parts of $A_0$, we can now work toward the physical kinematics used for the calculation of $A_2$ described above.  The same Iwasaki plus DSDR gauge action but with the sea quark mass reduced to its physical value should permit calculation with physical pion and kaon masses. Recent RBC/UKQCD experience~\cite{Yin:2012xx} with the use of the Mobius variant of DWF~\cite{Brower:2004xi, Brower:2012vk}  suggests that the extent in the fifth dimension of 32 used in the current Iwasaki + DSDR, DWF calculations can be reduced to 16 to enable a high statistics study of $A_0$.  More challenging is the needed non-zero relative momentum of the two final-state pions.  As discussed above we plan to accomplish this by imposing G-parity boundary conditions on the sea and valence quarks.  This effort is well underway in a project carried out by C.~Kelly.  Quenched studies have been performed~\cite{Kelly:2012xx} and code written which is currently under test to generate $N_f=2+1$ gauge configurations obeying G-parity boundary conditions.  

While a large reduction in the errors from the disconnected diagrams results from the use of split-pion, Coulomb gauge fixed wall sources, we are optimistic that all-to-all propagator methods~\cite{Foley:2005ac,Aoki:2009qn} will allow us to construct even more effective split-pion sources using localized pion wave functions, possibly gaining a further factor of two reduction in statistical error.  A final critical acceleration is provided by the advance of computer technology.  With the now available BG/Q computer hardware and highly efficient QCD code described in Peter Boyle's talk at this meeting~\cite{Boyle:2012zz}, the generation of 3K time units of a specialized $32^3 \times 64$ G-parity ensemble requires only a few months on a BG/Q 1024-node rack.  The difficult, high-statistics measurements used to obtain the $24^3\times 64$ results for $A_0$ presented here formed the final project carried out on a 4096-node QCDOC partition in 2011.  The sustained performance and memory size of this partition (1 Tflops/500 Gbytes) are equaled by 32 BG/Q nodes, allowing a calculation on a 512 node BG/Q partition to run 16 times faster.  While important aspects of a calculation of $A_0$ (which will give the standard model prediction for $\epsilon'/\epsilon$) with physical kinematics will remain uncertain until large-scale experiments are begun in a few months, this calculation should be possible with present resources.

\section{Computing the $K_L - K_S$ mass difference}

Much must yet be accomplished to accurately carry out the $K\to\pi\pi$ decay calculations discussed in the previous section.  However, it is plausible that the basic methods are now understood and that presently available measurement algorithms and computer resources will be sufficient for the task.  In this last section of the talk, we will discuss a more difficult topic at a much earlier stage in development, the calculation of ``long distance'' contributions to second order weak processes.  One might expect a large difference in complexity between the first order (one-$W$ exchange) processes responsible for $K\to\pi\pi$ decay and the second order weak  (two-$W$ exchange) processes needed for $K^0 - \overline{K^0}$ mixing.  However, if the two-$W$ exchange is dominated by momenta on the order of the $W$ boson or top quark mass, then at the scale of hadronic phenomena this second order process will appear to take place at a point and can be represented by a four-quark operator in a lattice QCD calculation just as is done for processes that involve a single $W$ exchange.  

\begin{figure}[t]
\centering
\includegraphics*[width=0.8\textwidth]{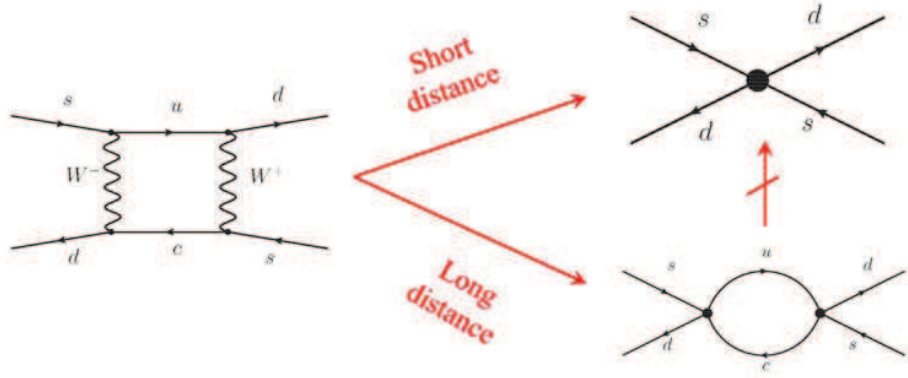}
\caption{\label{fig:distances} The two possible low energy descriptions of a box diagram entering $K^0 - \overline{K^0}$ mixing. Note there can be overlap between these long- and short-distance contributions.  The long-distance part, expressed as the product of two weak operators $H_W$, will itself contain a short distance part when these two operators collide in space-time.  Such a short distance component of this long distance amplitude will typically be incorrect but, if significant, can be removed by explicit subtraction.}
\end{figure}

The CP violating, $K^0 - \overline{K^0}$ mixing amplitude is dominated by short distance effects and the corresponding mixing parameter $\epsilon_K$ is typically expressed as a product of a short distance Wilson coefficient, derived from a box diagram in which two $W$ bosons are exchanged, and the low energy matrix element of a four-quark operator evaluated between $K^0$ and $ \overline{K^0}$ states, which determines the familiar $B_K$ parameter.  However, there is also an $\approx 5\%$ ``long distance'' contribution to  $\epsilon_K$~\cite{Buras:2010pza} in which the two exchanged $W$ mesons are separated by distances on the order of $1/\Lambda_{\rm QCD}$ or $1/m_\pi$.   Further, the CP conserving part of the $K^0 - \overline{K^0}$ mixing amplitude which gives the $K_L-K_S$ mass difference $\Delta M_K$,  receives a potentially large, long distance contribution.  One expects that the largest contribution to $\Delta M_K$ comes from distances on the order of the inverse charm quark mass, $1/m_c$.  This is conventionally referred to as a short distance contribution.  However, perturbation theory calculations show large NNLO terms~\cite{Brod:2010mj} suggesting that this scale might be better thought of as also a relatively long distance at which perturbation theory has become unreliable and again a lattice QCD calculation is needed.  These short- and long-distance contributions to a box diagram which enters $K^0 - \overline{K^0}$ mixing are illustrated in Fig.~\ref{fig:distances}.

Two important questions arise when considering a lattice QCD calculation of long distance effects in second-order weak amplitudes.  The first is associated with the use of Euclidean space methods to compute amplitudes with possible real intermediate states.  These states result in principal parts appearing in formula for the real parts and give imaginary parts which cannot appear in a Euclidean space calculation.  The second issue is the need to develop effective lattice methods to evaluate the resulting four-point functions in which a $K^0$ is transformed into a $\overline{K^0}$ by the action of two, separated, first order weak operators $H_W$.  We will now consider each of these issues in turn.

The computation of long distance parts of second order weak processes is key to many important weak processes, for example rare $K$ decays~\cite{Isidori:2005tv}.  Here we will focus here on $K^0 - \overline{K^0}$ mixing, typically described in the Wigner-Weisskopf formalism by the evolution equation~\cite{Donoghue:1992dd}:
\begin{equation}
i\frac{d}{dt}\left(\begin{array}{c} K^0 \\ \overline{K}^0 \end{array}\right) 
= \left\{ \left( \begin{array}{cc} M_{00} & M_{0\overline{0}} \\
                        M_{\overline{0}0} & M_{\overline{0} \overline{0}}
                           \end{array} \right) 
- \frac{i}{2} \left( \begin{array}{cc} \Gamma_{00} & \Gamma_{0\overline{0}} \\
               \Gamma_{\overline{0}0} &\Gamma_{\overline{0} \overline{0}}
                           \end{array} \right)\right\} 
\left(\begin{array}{c} K^0 \\ \overline{K}^0 \end{array}\right)
\label{eq:WW}
\end{equation}
where the $2 \times 2$ matrices $M$ and $\Gamma$ are given by:
\begin{eqnarray}
\Gamma_{ij} &=& 2\pi \sum_\alpha \int_{2m_\pi}^\infty d E
                      \langle i |H_W|\alpha(E)\rangle
                      \langle \alpha(E)|H_W|j\rangle \delta(E-m_K)  
\label{eq:WW-Gamma} \\
M_{ij} &=& \sum_\alpha {\cal P} \int_{2m_\pi}^\infty d E \frac{
                       \langle i |H_W|\alpha(E)\rangle
                       \langle \alpha(E)|H_W|j\rangle}{m_K - E}. 
\label{eq:WW-M}
\end{eqnarray}
We are using the subscripts $0$ and $\overline{0}$ to represent the $K^0$ and $\overline{K^0}$ states and the generalized sum over $\alpha$ and integral over the energy represents the sum over a complete set of energy eigenstates.

In the method of Lellouch and Luscher, the Luscher finite volume condition~\cite{Luscher:1990ux} is used to relate the finite volume energy of the $K- \pi\pi$ system to the infinite volume $\pi-\pi$ scattering phase shift evaluated at the resonant energy corresponding to on-shell $K\to\pi\pi$ decay.  Both quantities are evaluated to first order in $H_W$: the first order shift in the finite volume $\pi-\pi$ energy coming from the coupling to the degenerate $K$ state and the Breit-Wigner resonant contribution to $\pi-\pi$ scattering evaluated at the kaon pole.  

A generalization of this approach connects the infinite-volume, second-order mass difference $\Delta M_K$ which shifts the location of the kaon pole in resonant $\pi-\pi$ scattering to the  second order degenerate perturbation theory of a finite volume system including two or three nearly degenerate states: a two-pion state and either a CP even or both CP even and odd $K^0 - \overline{K}^0$ states, depending on whether we wish to treat only $\Delta M_K$ or both $\Delta M_K$ and $\epsilon_K$~\cite{Christ:2010zz,Christ:2012np}.  In contrast to the Lellouch-Luscher correction, the effect of finite volume is not a multiplicative factor but an additive, $1/L^3$, correction.  For $\Delta M_K$ the infinite- and finite-volume expressions are related through $O(1/L^3)$ by
\begin{eqnarray}
\Delta M_K  &=& 
2\sum_{n \ne n_0}\frac{\langle\overline{K^0}|H_W|n\rangle
                                 \langle n|H_W|K^0\rangle}{m_K-E_n}
 +\frac{1}{\frac{\partial (\phi+\delta_{\,0})}{\partial E}}
  \Bigg[\frac{1}{2}\frac{\partial^2 (\phi+\delta_{\,0})}{\partial E^2} 
      |\langle n_0|H_W|K_S\rangle|^2 \nonumber \\
&& -\frac{\partial}{\partial E_{n_0}}\left\{
    \left.\frac{\partial(\phi+\delta_{\,0})}{\partial E}\right|_{E=E_{n_0}}\hskip -0.1in 
    |\langle n|H_W|K_S\rangle|^2\right\}\Bigg] .
\label{eq:result}
\end{eqnarray}
Here the state $|n_0\rangle$ is a two-pion state whose energy has been adjusted to equal that of the $K$ meson by the choice of volume.  This result suggests that just as for the $K\to\pi\pi$ decay amplitude, the long distance parts of second-order weak $K^0 - \overline{K^0}$ mixing amplitude should be accessible to a Euclidean space calculation with controlled finite-volume errors.  

\begin{figure}[t]
\centering
\includegraphics*[width=0.8\textwidth]{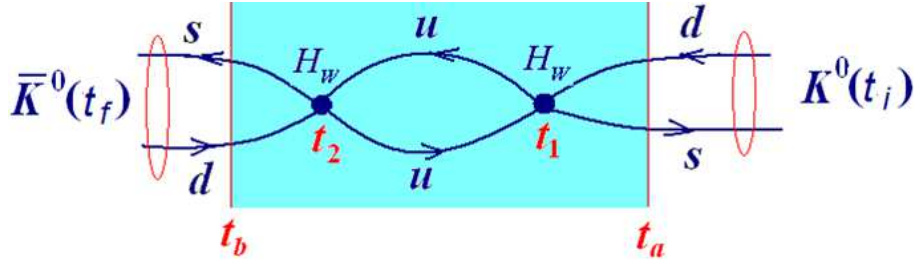}
\caption{\label{fig:DeltaM_K_amp}  A diagram showing the elements of a lattice calculation of the long distance contribution to $\Delta M_K$.  The times $t_1$ and $t_2$ appearing in the two effective weak operators are integrated over a fixed interval $t_a \le t_2,t_1 \le t_b$.  In Minkowski space such an amplitude would yield the second order mass shift times the elapsed time interval $t_b-t_a$.}
\end{figure}

We now discuss the methods needed to carry out such a second-order weak calculation using lattice QCD.  This has been an active project of the RBC/UKQCD collaboration for the past two years and has been carried out and reported on at both the current and previous Lattice Field Theory Symposia by J. Yu~\cite{Yu:2011gk,Yu:2012xx}.  The basic idea is to integrate the product of two effective weak Hamiltonia $H_W(t_i)$, $i=1,2$ over a fixed temporal region $t_a \le t_1,t_2 \le t_b$, taking the matrix element between $K^0$ and $\overline{K^0}$ states:
\begin{equation}
\mathcal{A} = \langle 0|T\left(K^0(t_f)\frac{1}{2}\int_{t_a}^{t_b}dt_2\int_{t_a}^{t_b}dt_1 H_W(t_2) H_W(t_1)\overline{K^0}(t_i)\right)|0\rangle.
\label{eq:DeltaM_K_1}
\end{equation}
This product is illustrated in Fig.~\ref{fig:DeltaM_K_amp}.

We can evaluate Eq.~\eqref{eq:DeltaM_K_1} by inserting sums over complete sets of intermediate states and carrying out the integrals over $t_1$ and $t_2$.  These are actually sums over discrete times which can be evaluated as partial geometric series and approximated for small $a$ to give:
\begin{eqnarray}
\mathcal{A}&=&N_K^2 e^{-M_K(t_f-t_i)}\Biggl\{\sum_{n\ne n_0} 
 \frac{\langle\overline{K^0}|H_W|n\rangle
         \langle n|H_W|K^0\rangle}{M_K-E_n}
 \Biggl(-(t_b-t_a) - \frac{1}{M_K-E_n} \nonumber \\
&&\hskip 0.5 in+\frac{e^{(M_K-E_n)(t_b-t_a)}}{M_K-E_n}\Biggr) 
 + \frac{1}{2}\langle\overline{K^0}|H_W|n_0\rangle
         \langle n_0|H_W|K^0\rangle(t_b-t_a)^2\Biggr\}
\label{eq:DeltaM_K_2}.
\end{eqnarray}
This equation contains four terms which should be interpreted: three within the large curved brackets and the fourth term with the 1/2 prefactor.  The first term, proportional to $t_b-t_a$, is the desired finite-volume expression of the mass difference.  The second is an uninteresting, time-independent constant, the third disappears at large time for all states more massive that the kaon.  States which are lighter than the kaon will give exponentially growing contributions which must be evaluated separately and subtracted.  The fourth term, proportional to $(t_b-t_a)^2$, results if the volume has been adjusted to create a state, $|n_0\rangle$, degenerate with the $K$.  This term must also be identified and discarded if the prescription to control finite volume errors described above is followed.

\begin{figure}[t]
\centering
\subfigure[Type 1]{\includegraphics*[width=0.37\textwidth]{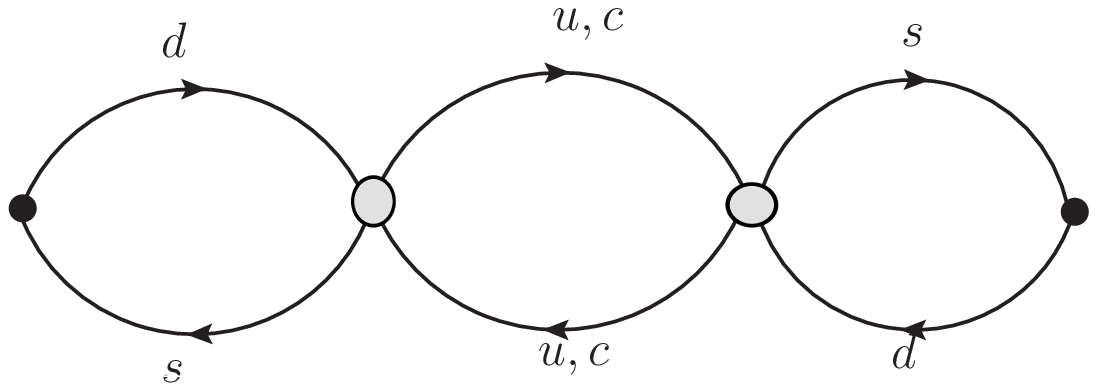}} \hskip 0.4in
\subfigure[Type 2]{\includegraphics*[width=0.37\textwidth]{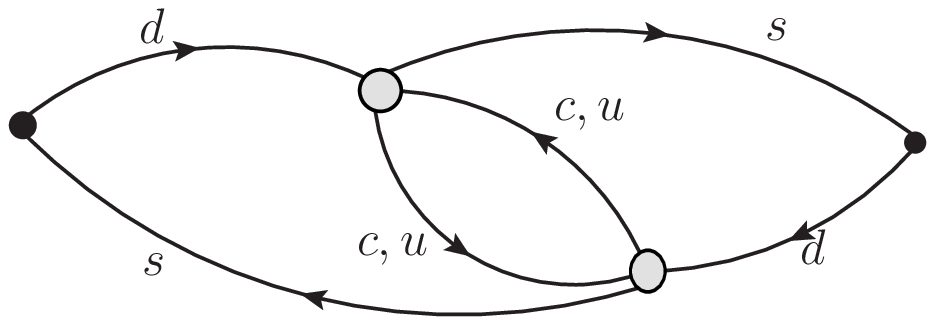}} 
\subfigure[Type 3]{\includegraphics*[width=0.37\textwidth]{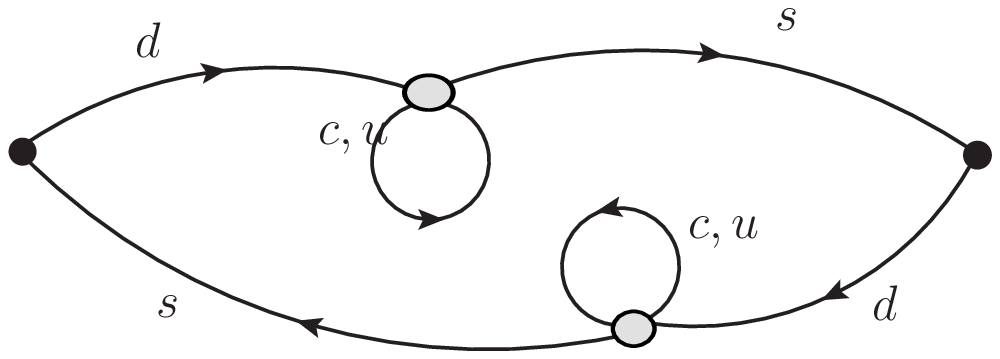}} \hskip 0.4in
\subfigure[Type 4]{\includegraphics*[width=0.37\textwidth]{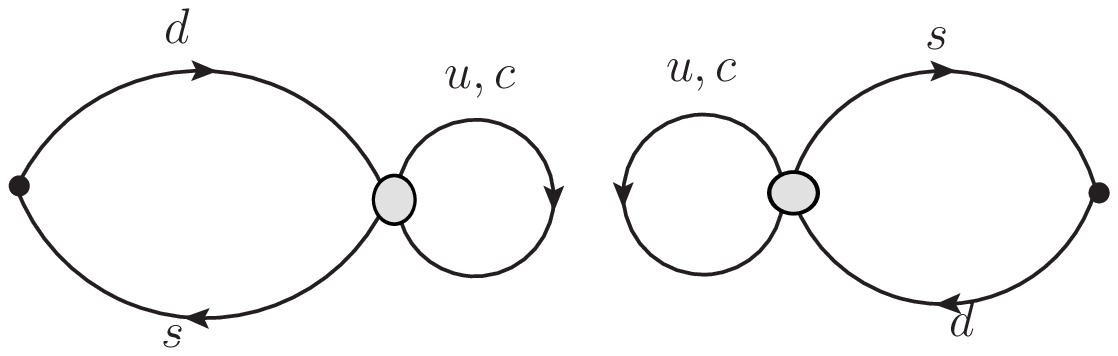}}
\caption{\label{fig:Types_DeltaM_K} Examples of the four types of diagram that enter the calculation of the second order weak contribution to $\Delta M_K$.}
\end{figure}

Figure~\ref{fig:Types_DeltaM_K} shows the four types of diagram which contribute to the amplitude $\mathcal{A}$.  In the first exploratory calculation of this quantity~\cite{Yu:2011gk,Yu:2012xx}, diagrams of type 3 and 4 which are disconnected in the $t$ or $s$ channel are neglected.  This calculation is performed on an ensemble of $16^3\times 32$ configurations with $1/a = 1.73$ GeV, generated using 2+1 flavors and the Iwasaki gauge action.   In order to realize GIM cancellation, a valence charm quark is included and the effective weak Hamiltonian appropriate for four flavors is used.  This Hamiltonian includes the six weak operators
\begin{equation}
Q_1^{qq'} = \overline{s_i}\gamma^\mu(1-\gamma^5)d_i \overline{q}_j\gamma^\mu(1-\gamma^5)q'_j
\quad Q_2^{qq'}  = \overline{s}_i\gamma^\mu(1-\gamma^5)d_j \overline{q}_j\gamma^\mu(1-\gamma^5)q'_i
\end{equation}
where $i$ and $j$ are color indices while $q$ and $q'$ are $u$ and/or $c$ quarks.  The introduction of a charm quark into a lattice calculation with an inverse lattice spacing $1/a= 1.73$ GeV will introduce potentially large discretization errors requiring future work with smaller $a$.  The GIM cancellation is complete: the short distance part of our $H_W \times H_W$ product is inaccurate at the level of $(m_c/m_W)^2$, much smaller than the $(m_ca)^2$ discretization errors.

\begin{figure}[t]
\centering
\subfigure[The integrated correlator $\mathcal{A}$ as a function of the integration interval $t_b-t_a$ after GIM cancellation with a 0.954 GeV valence charm quark. The red squares and blue diamonds are the results before and after the subtraction of the exponentially increasing $\pi^0$ term respectively. We include only the $Q_1 \cdot Q_1$ operator combination in this plot.]
{\includegraphics*[width=0.4\textwidth]{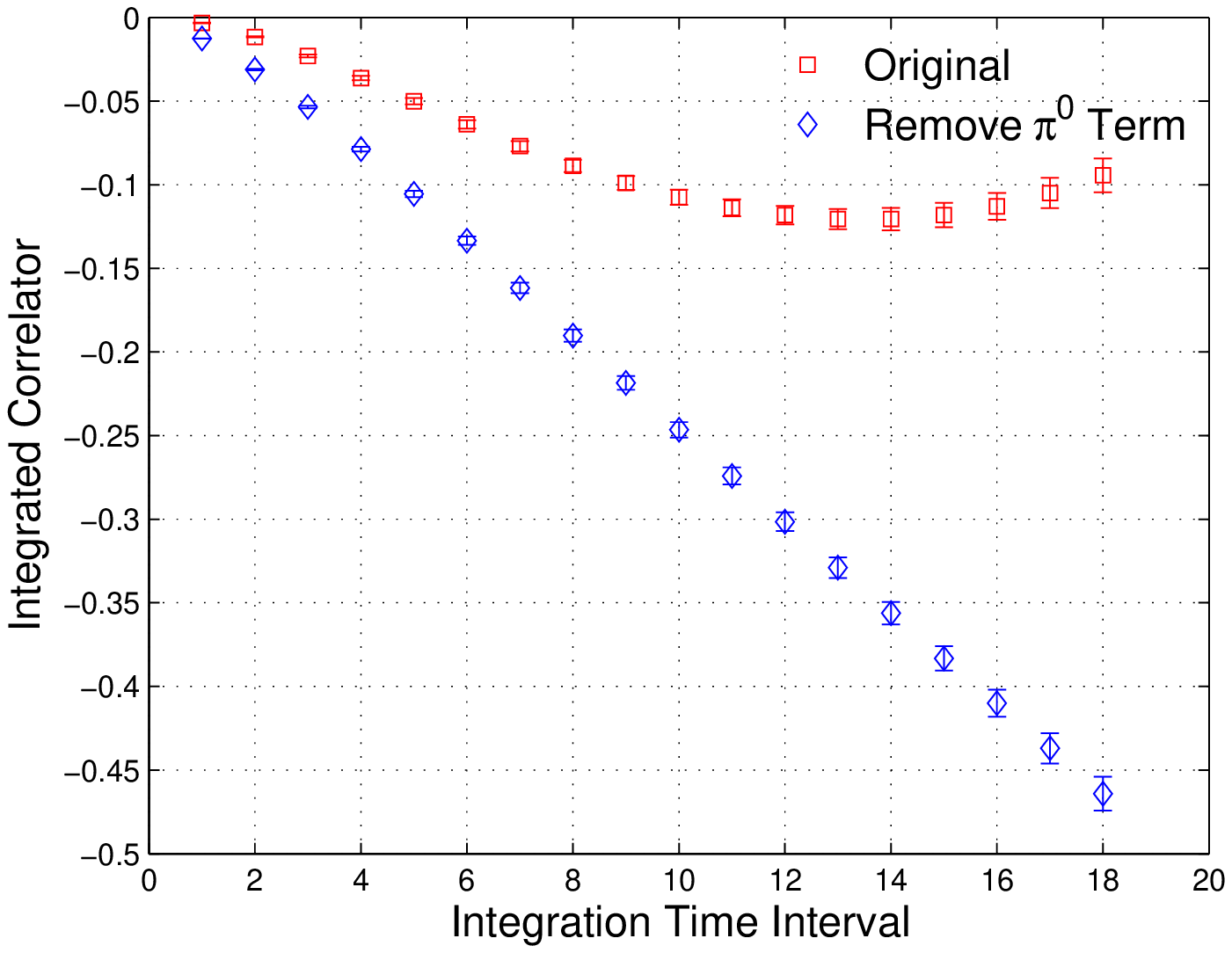}}
\hskip 0.2in
\subfigure[The mass difference $\Delta M_K^{11}$, obtained from the slope of the amplitude in Eq.~(\protect\ref{eq:DeltaM_K_2}) with respect to $t_b-t_a$ after GIM cancellation and  subtraction of the light pion contribution, plotted as a function of the valence charm quark mass.  The `11' superscipt indicates that only the contribution of the product of operators of the type $Q_1$ is shown.]
{\includegraphics*[width=0.4\textwidth]{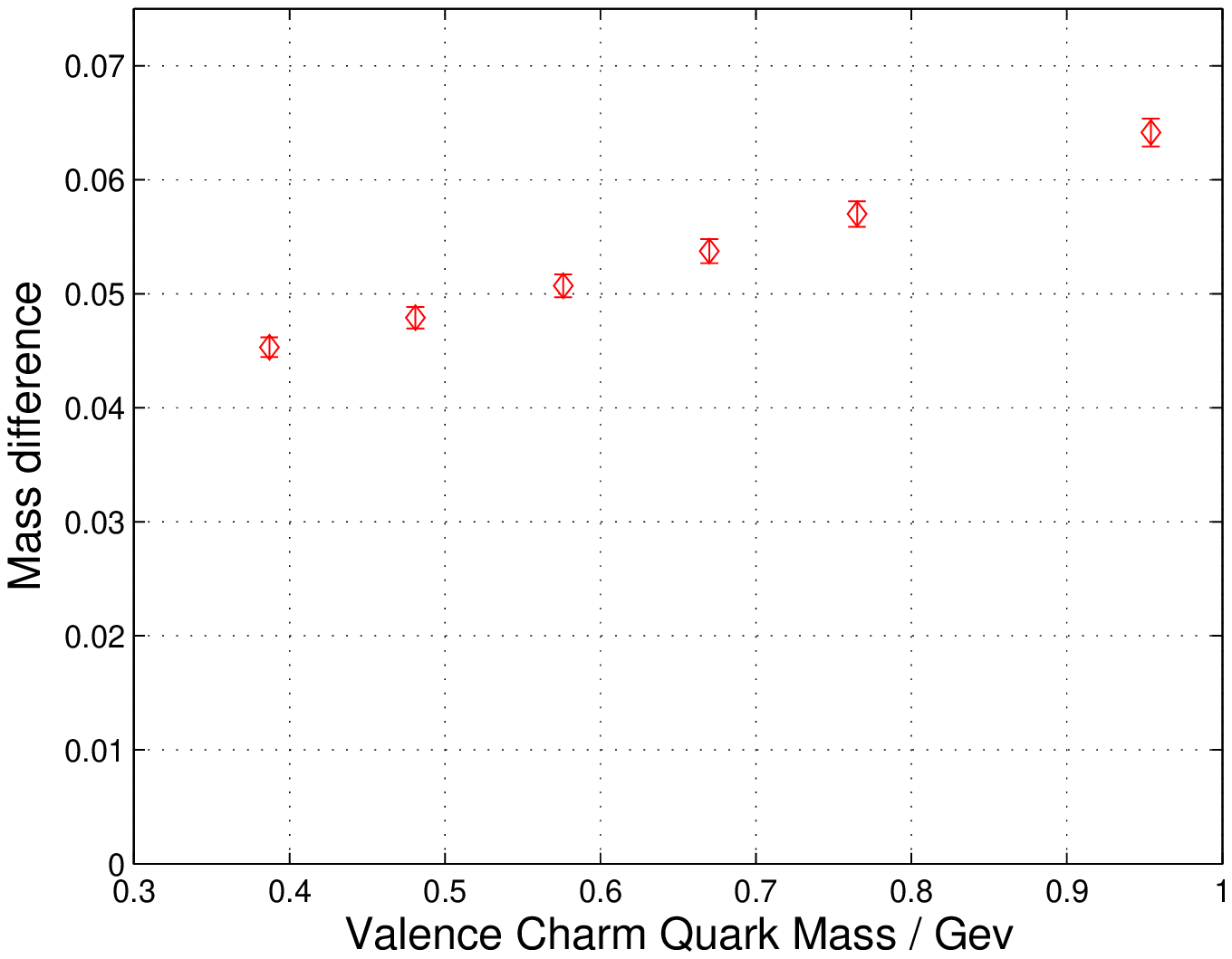}}
\caption{\label{fig:DeltaM_K_results} Results from the lattice QCD calculation of $\Delta M_K$.}
\end{figure}

Preliminary results are shown in Fig.~\ref{fig:DeltaM_K_results}.  The left panel shows the linear behavior as the integration time interval $t_b-t_a$ is varied suggesting that the required slope is not difficult to extract, after the exponentially growing contribution from the light $\pi^0$ state has been removed.  The right panel shows the increasing values for $\Delta M_K$ that result for increasing charm quark mass, $m_c$.  This figure also suggests the presence of a sizable constant term needed to describe the large $m_c$ dependence, reflecting a significant, $m_c$-independent long-distance contribution, at least for the large values of light and strange quark masses used here.   The final results for $\Delta M_K$ vary between $5.12(24)$ and $9.31(65) \; 10^{-12}$ MeV as the kaon mass varies between 563 and 834 MeV, for a pion mass of 421 MeV.  These results are somewhat larger than the experimental value of $3.483(6)\;10^{-12}$ MeV.

\section{Conclusion}

Substantial advances in computer capability and powerful new numerical methods are dramatically increasing the accuracy with which standard quantities can be computed using lattice QCD and expanding the range of important quantities which can be calculated.  By working at physical light quark mass and relatively coarse lattice spacing, it is now possible to directly calculate the $I=2$ $K\to\pi\pi$ decay amplitude $A_2$ with the largest error coming from finite lattice spacing effects.  Over the next couple of years, it should be possible to repeat these calculations on a series of ensembles with varying lattice spacing, reducing the error on $A_2$, perhaps to the 5\% level expected from isospin breaking effects.  The more difficult $I=0$ amplitude $A_0$ has been computed for unphysical kinematics and calculations with physical kinematics are now being actively planned.  Much more ambitious is the calculation of the $K_L-K_S$ mass difference $\Delta M_K$ which appears to be within reach.  While exploratory calculations are now underway, results with controlled errors on the 5\% level for $\Delta M_K$ are likely five years away.  The results reported here represent a major research direction of the RBC and UKQCD collaborations and I thank my collaborators whose work is being described.

\bibliography{kpipi}

\providecommand{\href}[2]{#2}\begingroup\raggedright\begin{thebibliography}{10}

\bibitem{Blum:2011ng}
T.~Blum, P.~Boyle, N.~Christ, N.~Garron, E.~Goode {\em et~al.}, {\it {The
  $K\to(\pi\pi)_{I=2}$ Decay Amplitude from Lattice QCD}},  {\em
  Phys.Rev.Lett.} {\bf 108} (2012) 141601
  [\href{http://arXiv.org/abs/1111.1699}{{\tt arXiv:1111.1699 [hep-lat]}}].
%%CITATION = ARXIV:1111.1699;%%

\bibitem{Blum:2012uk}
T.~Blum, P.~Boyle, N.~Christ, N.~Garron, E.~Goode {\em et~al.}, {\it {Lattice
  determination of the $K \to (\pi\pi)_{I=2}$ Decay Amplitude $A_2$}},  {\em
  Phys.Rev.} {\bf D86} (2012) 074513
  [\href{http://arXiv.org/abs/1206.5142}{{\tt arXiv:1206.5142 [hep-lat]}}].
%%CITATION = ARXIV:1206.5142;%%

\bibitem{Buchalla:1995vs}
G.~Buchalla, A.~J. Buras and M.~E. Lautenbacher, {\it Weak decays beyond
  leading logarithms},  {\em Rev. Mod. Phys.} {\bf 68} (1996) 1125--1144
  [\href{http://arXiv.org/abs/hep-ph/9512380}{{\tt hep-ph/9512380}}].
%%CITATION = HEP-PH 9512380;%%

\bibitem{Blum:2001xb}
{\bf RBC} Collaboration, T.~Blum {\em et~al.}, {\it Kaon matrix elements and
  cp-violation from quenched lattice qcd. i: The 3-flavor case},  {\em Phys.
  Rev.} {\bf D68} (2003) 114506
  [\href{http://arXiv.org/abs/hep-lat/0110075}{{\tt hep-lat/0110075}}].
%%CITATION = HEP-LAT 0110075;%%

\bibitem{Martinelli:1995ty}
G.~Martinelli, C.~Pittori, C.~T. Sachrajda, M.~Testa and A.~Vladikas, {\it A
  general method for nonperturbative renormalization of lattice operators},
  {\em Nucl. Phys.} {\bf B445} (1995) 81--108
  [\href{http://arXiv.org/abs/hep-lat/9411010}{{\tt hep-lat/9411010}}].
%%CITATION = NUPHA,B445,81;%%

\bibitem{Aoki:2007xm}
Y.~Aoki {\em et~al.}, {\it {Non-perturbative renormalization of quark bilinear
  operators and $B_K$ using domain wall fermions}},  {\em Phys. Rev.} {\bf D78}
  (2008) 054510 [\href{http://arXiv.org/abs/0712.1061}{{\tt arXiv:0712.1061
  [hep-lat]}}].
%%CITATION = 0712.1061;%%

\bibitem{Arthur:2010ht}
{\bf RBC Collaboration, UKQCD Collaboration} Collaboration, R.~Arthur and
  P.~Boyle, {\it {Step Scaling with off-shell renormalisation}},  {\em
  Phys.Rev.} {\bf D83} (2011) 114511
  [\href{http://arXiv.org/abs/1006.0422}{{\tt arXiv:1006.0422 [hep-lat]}}].
%%CITATION = ARXIV:1006.0422;%%

\bibitem{Arthur:2011cn}
{\bf RBC and UKQCD Collaborations} Collaboration, R.~Arthur, P.~Boyle,
  N.~Garron, C.~Kelly and A.~Lytle, {\it {Opening the Rome-Southampton window
  for operator mixing matrices}},  {\em Phys.Rev.} {\bf D85} (2012) 014501
  [\href{http://arXiv.org/abs/1109.1223}{{\tt arXiv:1109.1223 [hep-lat]}}].
%%CITATION = ARXIV:1109.1223;%%

\bibitem{Lellouch:2000pv}
L.~Lellouch and M.~Luscher, {\it Weak transition matrix elements from
  finite-volume correlation functions},  {\em Commun. Math. Phys.} {\bf 219}
  (2001) 31--44 [\href{http://arXiv.org/abs/hep-lat/0003023}{{\tt
  hep-lat/0003023}}].
%%CITATION = HEP-LAT 0003023;%%

\bibitem{Kim:2002np}
C.-h. Kim and N.~H. Christ, {\it K --> pi pi decay amplitudes from the
  lattice},  {\em Nucl. Phys. Proc. Suppl.} {\bf 119} (2003) 365--367
  [\href{http://arXiv.org/abs/hep-lat/0210003}{{\tt hep-lat/0210003}}].
%%CITATION = HEP-LAT 0210003;%%

\bibitem{Kim:2003xt}
C.~Kim, {\it {I = 2 pi pi scattering using G-parity boundary condition}},  {\em
  Nucl. Phys. Proc. Suppl.} {\bf 129} (2004) 197--199
  [\href{http://arXiv.org/abs/hep-lat/0311003}{{\tt arXiv:hep-lat/0311003}}].
%%CITATION = HEP-LAT/0311003;%%

\bibitem{Kim:2009fe}
C.~Kim and N.~H. Christ, {\it {G parity boundary conditions and Delta I = 1/2,
  K to pi pi decays}},  {\em PoS} {\bf LAT2009} (2009) 255
  [\href{http://arXiv.org/abs/0912.2936}{{\tt arXiv:0912.2936 [hep-lat]}}].
%%CITATION = 0912.2936;%%

\bibitem{Sachrajda:2004mi}
C.~T. Sachrajda and G.~Villadoro, {\it {Twisted boundary conditions in lattice
  simulations}},  {\em Phys. Lett.} {\bf B609} (2005) 73--85
  [\href{http://arXiv.org/abs/hep-lat/0411033}{{\tt arXiv:hep-lat/0411033}}].
%%CITATION = HEP-LAT/0411033;%%

\bibitem{Kim:2004sk}
C.~Kim, {\it Lattice calculation of delta isospin = 3/2 kaon decays to pion
  pion decay amplitude with interacting two pions}, . UMI-31-47246.

\bibitem{Yamazaki:2008hg}
{\bf RBC Collaboration, UKQCD Collaboration} Collaboration, T.~Yamazaki, {\it
  {On-shell Delta I = 3/2 kaon weak matrix elements with non-zero total
  momentum}},  {\em Phys.Rev.} {\bf D79} (2009) 094506
  [\href{http://arXiv.org/abs/0807.3130}{{\tt arXiv:0807.3130 [hep-lat]}}].
%%CITATION = ARXIV:0807.3130;%%

\bibitem{Hoogland:1977kt}
W.~Hoogland, S.~Peters, G.~Grayer, B.~Hyams, P.~Weilhammer {\em et~al.}, {\it
  {Measurement and Analysis of the pi+ pi+ System Produced at Small Momentum
  Transfer in the Reaction pi+ p- $\to$ pi+ pi+ n at 12.5-GeV}},  {\em
  Nucl.Phys.} {\bf B126} (1977) 109.
%%CITATION = NUPHA,B126,109;%%

\bibitem{Losty:1973et}
M.~Losty, V.~Chaloupka, A.~Ferrando, L.~Montanet, E.~Paul {\em et~al.}, {\it {A
  Study of pi-pi-scattering from pi-p interactions at 3.93-GeV/c}},  {\em
  Nucl.Phys.} {\bf B69} (1974) 185--204.
%%CITATION = NUPHA,B69,185;%%

\bibitem{Schenk:1991xe}
A.~Schenk, {\it {Absorption and dispersion of pions at finite temperature}},
  {\em Nucl.Phys.} {\bf B363} (1991) 97--116.
%%CITATION = NUPHA,B363,97;%%

\bibitem{Luscher:1990ux}
M.~Luscher, {\it Two particle states on a torus and their relation to the
  scattering matrix},  {\em Nucl. Phys.} {\bf B354} (1991) 531--578.
%%CITATION = NUPHA,B354,531;%%

\bibitem{Blum:2011pu}
T.~Blum, P.~Boyle, N.~Christ, N.~Garron, E.~Goode {\em et~al.}, {\it {$K$ to
  $\pi\pi$ Decay amplitudes from Lattice QCD}},  {\em Phys.Rev.} {\bf D84}
  (2011) 114503 [\href{http://arXiv.org/abs/1106.2714}{{\tt arXiv:1106.2714
  [hep-lat]}}].
%%CITATION = ARXIV:1106.2714;%%

\bibitem{Liu:2012xx}
Q.~Liu, {\it Kaon to two pions decays from lattice \uppercase{QCD}:
  $\uppercase{\Delta i}=1/2$ rule and \uppercase{CP} violation}, . ISBN:
  9781267290649.

\bibitem{Yin:2012xx}
{\bf RBC/UKQCD} Collaboration, H.~Yin, {\it Exploring qcd thermodynamics using
  moebius fermions},  {\em PoS} {\bf LAT2012} (2012) 191.

\bibitem{Brower:2004xi}
R.~C. Brower, H.~Neff and K.~Orginos, {\it Moebius fermions: Improved domain
  wall chiral fermions},  \href{http://arXiv.org/abs/hep-lat/0409118}{{\tt
  hep-lat/0409118}}.
%%CITATION = HEP-LAT 0409118;%%

\bibitem{Brower:2012vk}
R.~C. Brower, H.~Neff and K.~Orginos, {\it {The M\'obius Domain Wall Fermion
  Algorithm}},  \href{http://arXiv.org/abs/1206.5214}{{\tt arXiv:1206.5214
  [hep-lat]}}.
%%CITATION = ARXIV:1206.5214;%%

\bibitem{Kelly:2012xx}
{\bf RBC/UKQCD} Collaboration, C.~Kelly, {\it Progress towards
  $\uppercase{\Delta i} = 1/2$ $\uppercase{K}\to\pi\pi$ decays with
  \uppercase{G}-parity boundary conditions},  {\em PoS} {\bf LAT2012} (2012).

\bibitem{Foley:2005ac}
J.~Foley, K.~Jimmy~Juge, A.~O'Cais, M.~Peardon, S.~M. Ryan {\em et~al.}, {\it
  {Practical all-to-all propagators for lattice QCD}},  {\em
  Comput.Phys.Commun.} {\bf 172} (2005) 145--162
  [\href{http://arXiv.org/abs/hep-lat/0505023}{{\tt arXiv:hep-lat/0505023
  [hep-lat]}}].
%%CITATION = HEP-LAT/0505023;%%

\bibitem{Aoki:2009qn}
{\bf JLQCD Collaboration, TWQCD Collaboration} Collaboration, S.~Aoki {\em
  et~al.}, {\it {Pion form factors from two-flavor lattice QCD with exact
  chiral symmetry}},  {\em Phys.Rev.} {\bf D80} (2009) 034508
  [\href{http://arXiv.org/abs/0905.2465}{{\tt arXiv:0905.2465 [hep-lat]}}].
%%CITATION = ARXIV:0905.2465;%%

\bibitem{Boyle:2012zz}
P.~Boyle, {\it {The BlueGene/Q supercomputer}},  {\em PoS} {\bf LAT2012} 020.

\bibitem{Buras:2010pza}
A.~J. Buras, D.~Guadagnoli and G.~Isidori, {\it {On epsilon\_K beyond lowest
  order in the Operator Product Expansion}},  {\em Phys. Lett.} {\bf B688}
  (2010) 309--313 [\href{http://arXiv.org/abs/1002.3612}{{\tt arXiv:1002.3612
  [hep-ph]}}].
%%CITATION = 1002.3612;%%

\bibitem{Brod:2010mj}
J.~Brod and M.~Gorbahn, {\it {Epsilon\_K at Next-to-Next-to-Leading Order: The
  Charm-Top-Quark Contribution}},  {\em Phys.Rev.} {\bf D82} (2010) 094026
  [\href{http://arXiv.org/abs/1007.0684}{{\tt arXiv:1007.0684 [hep-ph]}}].
%%CITATION = ARXIV:1007.0684;%%

\bibitem{Isidori:2005tv}
G.~Isidori, G.~Martinelli and P.~Turchetti, {\it {Rare kaon decays on the
  lattice}},  {\em Phys.Lett.} {\bf B633} (2006) 75--83
  [\href{http://arXiv.org/abs/hep-lat/0506026}{{\tt arXiv:hep-lat/0506026
  [hep-lat]}}].
%%CITATION = HEP-LAT/0506026;%%

\bibitem{Donoghue:1992dd}
J.~F. Donoghue, E.~Golowich and B.~R. Holstein, {\it {Dynamics of the standard
  model}},  {\em Camb. Monogr. Part. Phys. Nucl. Phys. Cosmol.} {\bf 2} (1992)
  1--540.
%%CITATION = CMPCE,2,1;%%

\bibitem{Christ:2010zz}
{\bf RBC and UKQCD Collaborations} Collaboration, N.~H. Christ, {\it {Computing
  the long-distance contribution to second order weak amplitudes}},  {\em PoS}
  {\bf LATTICE2010} (2010) 300 [\href{http://arXiv.org/abs/1012.6034}{{\tt
  arXiv:1012.6034 [hep-lat]}}].

\bibitem{Christ:2012np}
N.~H. Christ, {\it {Computing the long-distance contribution to the kaon mixing
  parameter $\epsilon_K$}},  {\em PoS} {\bf LATTICE2011} (2011) 277
  [\href{http://arXiv.org/abs/1201.2065}{{\tt arXiv:1201.2065 [hep-lat]}}].
%%CITATION = ARXIV:1201.2065;%%

\bibitem{Yu:2011gk}
J.~Yu, {\it {Long distance contribution to $K_{L}$-$K_{S}$ mass difference}},
  {\em PoS} {\bf LATTICE2011} (2011) 297
  [\href{http://arXiv.org/abs/1111.6953}{{\tt arXiv:1111.6953 [hep-lat]}}].
%%CITATION = ARXIV:1111.6953;%%

\bibitem{Yu:2012xx}
{\bf RBC/UKQCD} Collaboration, J.~Yu, {\it Lattice calculation of the $k_l-k_s$
  mass difference},  {\em PoS} {\bf LAT2012} (2012) 129.

\end{thebibliography}\endgroup
\bibliographystyle{JHEP}

%\begin{thebibliography}{99}
%  \bibitem{...} ....
%\end{thebibliography}

\end{document}